\documentclass{article}
\usepackage[T1]{fontenc}
\usepackage[utf8]{inputenc}
\usepackage{ismir} 
\usepackage{amsmath,cite,url}
\usepackage{graphicx}
\usepackage{color}
\usepackage{booktabs}
\usepackage{kotex}
\usepackage{amssymb, amsfonts}
\usepackage{microtype} 
\usepackage{enumitem}

\title{Frame-Level Pansori Mode Classification with Complementary Audio Representations}

\multauthor
  {Sangheon Park$^1$ \hspace{1cm} Seonguk Ju$^2$ \hspace{1cm} Suin Chung$^2$}
  {{\bf Danbinaerin Han$^3$ \hspace{1cm} Dasaem Jeong$^2$}\\
  $^1$ School of Music, Georgia Institute of Technology, USA\\
  $^2$ Music \& Arts Learning (MALer) Lab, Sogang University, South Korea\\
  $^3$ Department of Culture Technology, KAIST, South Korea\\
  {\tt\small sangheon@gatech.edu, \{jinin5113, sorisori98, dasaemj\}@sogang.ac.kr, naerin71@kaist.ac.kr}
  }

\def\authorname{S. Park, S. Ju, S. Chung, D. Han, and D. Jeong}

\begin{document}

\setlength{\abovedisplayskip}{4pt plus 1pt minus 1pt}
\setlength{\belowdisplayskip}{4pt plus 1pt minus 1pt}
\setlength{\abovedisplayshortskip}{2pt plus 1pt minus 1pt}
\setlength{\belowdisplayshortskip}{2pt plus 1pt minus 1pt}

\maketitle

\begin{abstract}

Pansori is a traditional Korean vocal genre whose mode system (\emph{jo}) is defined not by scale alone but by the entanglement of pitch collection, microtonal ornament (\emph{sigimsae}), and vocal timbre. In this study, we introduce a 46-hour frame-level pansori mode annotation, expert-labeled across all five canonical \emph{batang}, and evaluate four complementary input representations (mel spectrogram, F0 contour, MIDI piano roll, and a multi-cultural SSL encoder) under two split strategies designed to detect shortcut learning. Across the three well-represented modes, performance degrades by only $2.1$--$3.6$ points of F1 when entire works are held out, indicating that the models learn mode-relevant features rather than memorizing repertoire. Per-class results further show that source separation removes the percussion cue on which \emph{changjo} depends, and that generic multi-cultural pre-training fails specifically on the Ujo--Gyemyeonjo distinction. Qualitative analysis of cross-modal disagreement recovers musicologically documented phenomena and agrees with published score-based analyses of modern \emph{changjak} pansori.

\end{abstract}

\section{Introduction}\label{sec:introduction}
The computational analysis of non-Western music traditions has become an increasingly active area within music information retrieval (MIR), with substantial progress on traditions such as Turkish, Hindustani, and Arabic art music, particularly in the study of their respective mode systems (\emph{makam}, \emph{raga}, and \emph{maqam})~\cite{borekci2024classification, clayton2022raga, alaydrus2023analysis}.
In each of these cases, computational approaches have helped formalize and probe musicological categories of traditional music. Yet \textit{pansori}, Korean traditional monologue storytelling, has been overlooked.

A \textit{pansori} mode(\textit{jo}, 조) is not adequately defined by a fixed scale or set of intervals~\cite{shin2018}. It is a \textit{multidimensional construct} encompassing pitch collections, microtonal ornaments (\textit{sigimsae}), idiomatic melodic patterns, vocal timbre, and an associated dramatic affect. 

Even within a single \textit{daemok} (narrative passage), the mode can shift several times in rapid succession. \textit{Ujo}, for example, is described as having a wide vocal range and a varied set 
of constituent pitches arranged in stable major-second intervals. Yet pansori mode is ultimately determined not by a do-re-mi-style scale but by the placement of \textit{yoseong} (vibrato) and \textit{kkeokneun-eum} (the bent note)~\cite{KimHyeJung1993}. Western staff notation, with its emphasis on discrete pitches and fixed intervals, cannot capture this entanglement of scale, ornamentation, and texture.

In this work, we approach pansori mode classification as a task that connects music information retrieval and ethnomusicology through a multi-representation framework.
Our contributions are fourfold: (i) a 46-hour annotated corpus of pansori recordings with frame-level mode labels across all five canonical \textit{batang}, annotated by a professional pansori singer; (ii) a multi-representation classification framework that operationalizes 
the theoretical dual definition of pansori mode—scale structure and performance style—through complementary input representations including MIDI piano roll, F0 contour, mel spectrogram, and pretrained music embeddings; (iii) a diagnostic evaluation protocol using Daemok-Shared Split and Work-level Split on a shared test set, demonstrating that our models learn mode features that generalize across singers and repertoire rather than memorizing song-specific content; and (iv) qualitative evidence that modality disagreement patterns recover the distinction between stylistic and scale-theoretic axes of mode, validated by expert listening and by agreement with published analyses of modern \emph{changjak} pansori.


\section{Background on Pansori Singing}\label{sec:background}

\subsection{Performance Tradition and Vocal Characteristics}

Pansori is a traditional Korean epic storytelling art, recognized in 2003 as a UNESCO Intangible Cultural Heritage of Humanity. A performance features a single vocalist (\textit{sorikkun}) who delivers a long-form narrative through a combination of singing (\textit{sori}), stylized speech (\textit{aniri}), and dramatic gesture (\textit{ballim}), accompanied only by a single barrel drum (\textit{buk}) played by a percussionist (\textit{gosu}). The pansori repertoire originally consisted of twelve epic narratives, of which only five \textit{batang} survive in performance today: \textit{Chunhyangga}, \textit{Simcheongga}, \textit{Heungboga}, \textit{Sugungga}, and \textit{Jeokbyeokga}.

A defining feature of pansori is its distinctive vocal production. Rather than the tonal purity prized in Western classical singing, pansori vocalists cultivate a raw, powerful acoustic texture—a forceful, unadulterated chest voice with a characteristically gravelly, husky timbre—through years of demanding physical training~\cite{LeeKyuHo1998}. This vocal texture is not a stylistic flourish but a load-bearing element of musical meaning: pansori modes are defined as much by their distinct vocal timbres as by their pitch material. The commanding \textit{Ujo} calls for an open, sonorous delivery, while the lamenting \textit{Gyemyeonjo} is realized through a constrained, narrowed vocal production. Acoustic texture is therefore indispensable for both human perception and computational analysis of pansori mode.

\subsection{Pansori Modes (\textit{Jo}) and Microtonality}


The musical architecture of pansori is governed by its mode system, \textit{jo} (조)~\cite{bohyung2012jo}. Two performances may share the same pitch set yet be classified as different modes on the basis of their ornamental and timbral characteristics alone. The repertoire is anchored by two primary modes---\textit{Gyemyeonjo} and \textit{Ujo}---with several minority modes that share characteristics with one or the other.

\begin{itemize}[noitemsep, topsep=0pt, parsep=0pt, partopsep=0pt, leftmargin=*]
    \item \textbf{\textit{Gyemyeonjo} (계면조).} Rooted in the folk and 
    shamanic ritual music of the southwestern Jeolla region, 
    \textit{Gyemyeonjo} is the most prevalent mode in pansori and is 
    used to express sorrow, lamentation, and pathos~\cite{KMS2_pansori}. 
    While structurally based on a pentatonic collection (typically 
    \textit{mi-la-do-re-mi}), its identity rests far more on 
    \textit{sigimsae} (microtonal ornaments) than on scale alone: a 
    wide, intensely vibrating lowest note (\textit{tteoneun-eum}), a 
    stable middle note (\textit{pyeong-eum}), and an upper note that 
    is sharply bent or smoothly glided downward 
    (\textit{kkeokneun-eum}). Because its essence lies in continuous 
    pitch trajectories rather than discrete notes, Gyemyeonjo 
    is particularly difficult to represent symbolically (e.g., in 
    MIDI) without losing what makes it distinctive.
    
    \item \textbf{\textit{Ujo} (우조).} Originating from aristocratic 
    vocal genres such as \textit{gagok} and \textit{sijo}, \textit{Ujo} 
    is associated with grandeur, heroism, and masculine majesty, and 
    typically accompanies depictions of noble characters or sweeping 
    scenery~\cite{KMS2_pansori}. It uses a \textit{sol-la-do-re-mi} 
    pentatonic scale and calls for a vigorous, commanding vocal 
    production originating from deep in the lower abdomen. In 
    contrast to Gyemyeonjo, pitches in Ujo are 
    relatively stable with broader and slower vibrato, melodic 
    intervals are wider, and the robust acoustic texture of the voice 
    is itself a defining hallmark of the mode.
\end{itemize}

In addition to these two sung modes, our annotation framework includes two non-modal categories that appear regularly in pansori performance: \textit{aniri}, the stylized speech-narration used to introduce dramatic context, deliver dialogue, and provide emotional relaxation between high-tension sung sections~\cite{KMS2_pansori}, and \textit{changjo}, an intermediate form between sung passages and \textit{aniri} that renders text in a free, improvisatory recitative-like melody at moments of heightened dramatic tension~\cite{KimJeongTae2008}. Although neither follows the canonical mode patterns of \textit{Gyemyeonjo} or \textit{Ujo}, both are integral to the musical fabric of a pansori performance and are annotated alongside the sung modes in our classification framework.

The remaining sub-modes—\textit{Pyeongjo}, \textit{Seollongje}, and \textit{Gyeongdeureum}—share \textit{Ujo}'s pentatonic scale structure and are conventionally grouped under the broader umbrella of \textit{Ujo}-adjacent singing in traditional pansori musicology~\cite{KMS2_pansori}. 

A direct consequence of this multidimensional definition is that no single representation can therefore characterize pansori mode: symbolic transcription captures the scalar foundation of \textit{Ujo} but discards the microtonality defining \textit{Gyemyeonjo}, while a spectrogram preserves timbre and continuous contour but blurs scale-theoretic structure. This motivates the multi-representation approach developed below.

\begin{table*}[t]
\centering
\small
\renewcommand{\arraystretch}{0.9}
\setlength{\tabcolsep}{5pt}
\begin{minipage}[t]{0.48\textwidth}
\centering
\begin{tabular}{lrr}
\toprule
\textbf{Label} & \textbf{Duration} & \textbf{\%} \\
\midrule
Gyemyeonjo (계면조) & 26h 45m & 58.2 \\
Ujo (우조)         & 8h 44m  & 19.0 \\
Aniri (아니리)     & 6h 59m  & 15.2 \\
Pyeongjo (평조)    & 2h 10m  & 4.7  \\
Changjo (창조)     & 50m     & 1.8  \\
Seollongje (설렁제)& 20m     & 0.8  \\
Gyeongdeureum (경드름) & 6m  & 0.2  \\
\midrule
\textbf{Total}     & \textbf{46h 02m} & \textbf{100.0} \\
\bottomrule
\end{tabular}
\end{minipage}
\hfill
\begin{minipage}[t]{0.48\textwidth}
\centering
\begin{tabular}{lrrr}
\toprule
\textbf{Category} & \textbf{Files} & \textbf{Duration} & \textbf{\%} \\
\midrule
Simcheongga (심청가) & 154 & 15h 36m & 33.9 \\
Chunhyangga (춘향가) & 74  & 11h 26m & 24.9 \\
Jeokbyeokga (적벽가) & 67  & 7h 53m  & 17.1 \\
Heungboga (흥보가)   & 40  & 4h 41m  & 10.2 \\
Sugungga (수궁가)    & 27  & 3h 34m  & 7.8  \\
Danga/Other (단가/기타) & 34 & 2h 51m & 6.2 \\
\multicolumn{4}{c}{} \\
\midrule
\textbf{Total}       & \textbf{396} & \textbf{46h 02m} & \textbf{100.0} \\
\bottomrule
\end{tabular}
\end{minipage}
\caption{Dataset composition: mode label distribution (left) and
song category (\emph{batang}) distribution (right).}
\label{tab:dataset}
\end{table*}

\section{Dataset}\label{dataset}

\subsection{Data Collection and Annotation}

The corpus comprises 396 tracks (46 h 02 m) spanning all five canonical batang together with danga and miscellaneous pieces, annotated at the frame level by a co-author holding a Ph.D. in Pansori Performance and Theory.
Each piece was segmented by mode transitions into seven standard pansori categories:
\textit{Gyemyeonjo}, \textit{Ujo}, \textit{Pyeongjo}, \textit{Aniri}, \textit{Changjo}, \textit{Seollongje}, \textit{Gyeongdeureum}.\footnote{Frame-level annotations, album metadata, and code are available at \url{https://github.com/michspark/Pansori-Mode-Classification}.}

For classification, we collapse the seven categories into four classes based on their musical and statistical properties: (i) \textit{Gyemyeonjo}, (ii) \textit{Ujo} (merging the minority modes \textit{Pyeongjo}, \textit{Seollongje}, and \textit{Gyeongdeureum}, each of which constitutes less than 5\% of the data), (iii) \textit{Aniri}, and (iv) \textit{Changjo}. The merger reflects established musicological convention: \textit{Pyeongjo} and \textit{Seollongje} share \textit{Ujo}'s \textit{sol-la-do-re-mi} pentatonic scale, differing primarily in register, mood, and dramatic function rather than in scalar identity, while \textit{Gyeongdeureum}, though derived from Seoul-Gyeonggi folk melodies, is conventionally grouped with \textit{Ujo}-adjacent singing for its bright, scale-based delivery that contrasts with the microtonal sadness of \textit{Gyemyeonjo}~\cite{KMS2_pansori}. This grouping thus preserves the principal scale-theoretic distinction in pansori—between \textit{Gyemyeonjo} and the \textit{Ujo}-family modes—while addressing the severe class imbalance that would otherwise preclude reliable evaluation of the minority modes. 

\subsection{Dataset Statistics}
The dataset is substantially imbalanced across both distributions
(Table~\ref{tab:dataset}), reflecting the natural prevalence of modes in the pansori repertoire.

\subsection{Splits}
Because of the limited number of \textit{daemok} in pansori, models risk learning passage-specific patterns rather than generalizing to unseen repertoire. To probe this risk, inspired by previous work on leitmotif detection, we construct two complementary split strategies~\cite{Krause-2021, 9054642}.

\textbf{Daemok-Shared Split.} This configuration evaluates the model's ability to generalize across singer-level variations while remaining within a known musical context. We deliberately selected 18 specific \textit{daemok} that have multiple recorded performances sung by different vocalists. For each \textit{daemok}, the performances were distributed across the training and test sets, ensuring the model is evaluated on the exact same passage but by an unseen singer. The remaining dataset, including the training portions of these 18 \textit{daemok}, was divided into training and validation sets at a 9:1 ratio. This split essentially asks: \textit{What is the performance upper bound when the model is already familiar with the underlying composition?}

\textbf{Work-level Split.} 
To evaluate true generalization to entirely unseen repertoire, entire \textit{daemok} are strictly partitioned so that no musical passage appears in both the training and test sets. Given the limited corpus of pansori---comprising only five canonical \textit{batang}---this is a particularly stringent test, ensuring models cannot rely on memorizing specific melodic contours or lyrics. We adopted a 10-fold cross-validation scheme to maximize evaluation robustness. Specifically, \textit{danga} (short introductory songs) were exclusively assigned to the training set. The five \textit{batang} were each divided into two segments, creating 10 distinct folds. In each iteration, 8 folds (covering 4 \textit{batang}) were used for training, while the remaining 2 folds (from the single held-out \textit{batang}) were split into validation and test sets. The performance gap between the two splits indicates whether models rely on passage-specific shortcuts versus learning intrinsic mode features.


\begin{figure*}[t]
    \centering
    \includegraphics[width=0.9\textwidth]{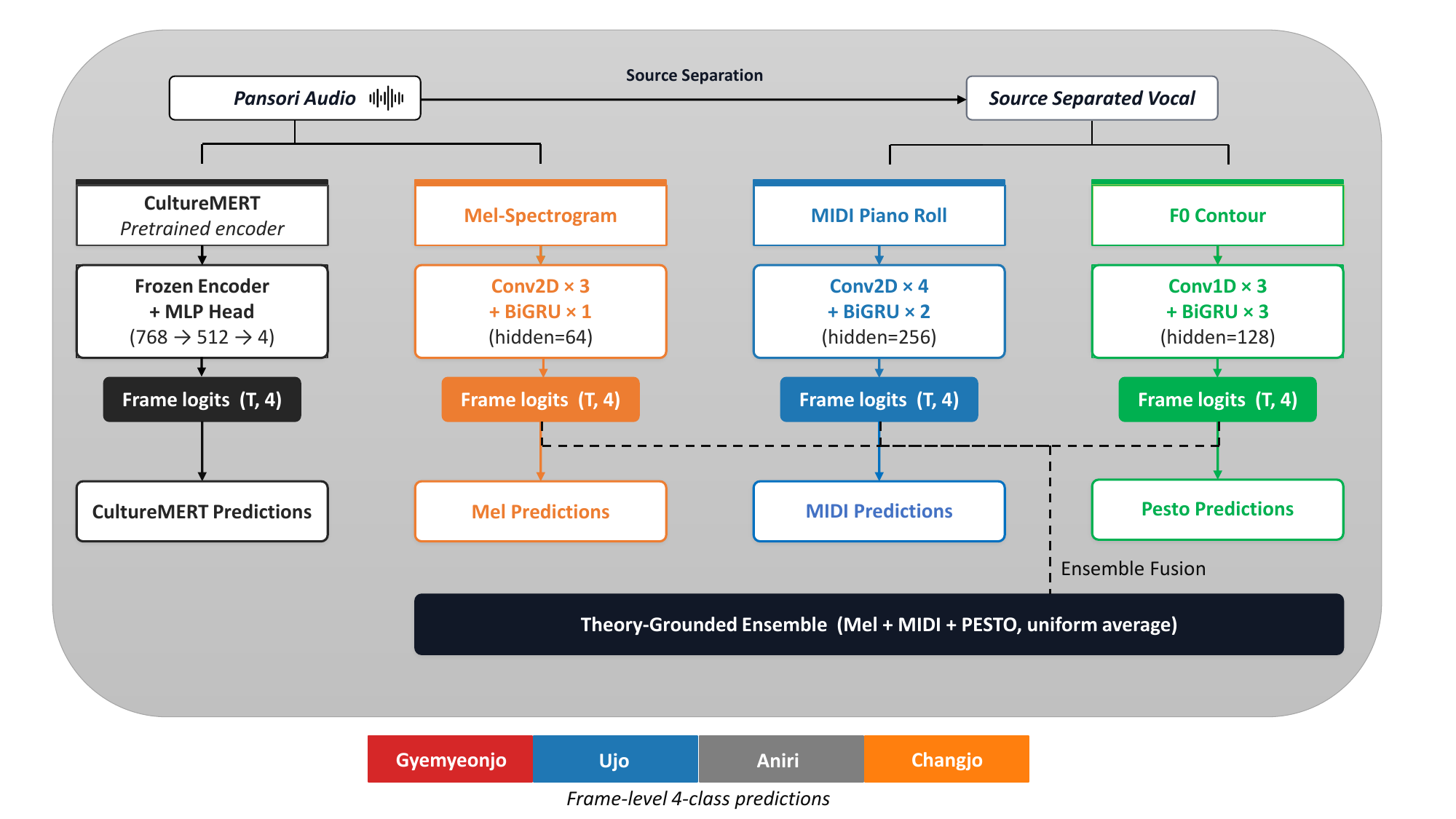}
    \caption{Multi-representation pansori mode classification framework.}
    \label{fig:framework}
\end{figure*}

\section{Pansori Mode Classification}\label{sec:method}
We formulate pansori mode classification as a frame-level sequence labeling task: given an audio recording, the model outputs a mode label for each frame. This formulation reflects the musicological reality that mode can shift within a single \emph{daemok}. At training and inference time, recordings are processed in fixed 30-second segments. All modalities follow a common CRNN-style design, which has been widely used for music classification tasks~\cite{choi2017, nasrullah2019}. This design has been successfully applied to comparable mode classification problems in non-Western traditions, including Hindustani raga identification~\cite{singh2024, madhusudhan2019deepsrgm}.

\subsection{Mel Spectrogram (Conv2DGRU)}
The mel spectrogram captures the overall spectral envelope, vocal timbre, and articulation, and has been the most widely adopted input representation for music classification tasks~\cite{choi2017, inbook}. We compute 128-band mel spectrograms at 16\,kHz (n\_fft\,=\,2048, hop\,=\,512) with log-amplitude normalization. To examine the effect of non-vocal components (drum, silence, ambient noise) on mode perception, we train models on both the original audio and source-separated vocals using Demucs~\cite{rouard2023hybrid}. The mel input is processed by a stack of three 2D convolutional blocks followed by a single-layer bidirectional GRU and a linear head producing per-frame logits. We apply audio-level perturbations (pitch shifting, Gaussian noise, gain scaling), time stretching applied jointly to features and labels, and SpecAugment-style spectral masking~\cite{park2019specaugment}.


\subsection{MIDI Piano Roll (Conv2DMIDI)}
The MIDI piano roll encodes music as a binary pitch-activation matrix over time and has been used effectively for symbolic music classification, including composer and style identification~\cite{kong2020largescalemidibasedcomposerclassification, yang2021composer}. After source separation with Demucs~\cite{rouard2023hybrid}, singing voice transcription is performed using the model of Li et al.~\cite{li2024}, and the resulting note events are quantized into a piano roll at 10\,fps over a 128-pitch-bin grid. To accommodate the wide pitch range and sparse, binary nature of this input, we deepen the convolutional stack relative to the mel-based model~\cite{kong2020largescalemidibasedcomposerclassification, yang2021composer}, followed by a 2-layer Bi-GRU. Training augmentation includes pitch shift and random time masking for robustness to transcription errors and performance variability.

\subsection{F0 Contour (Conv1DGRU)}
The F0 contour captures micro-pitch phenomena central to pansori performance practice---vibrato, bent notes, and gliding tones---and has been widely utilized in prior computational analyses of non-Western music~\cite{han2023finding, muluneh2024computational}. We extract continuous fundamental frequency estimates with PESTO~\cite{riou2025pesto}, filter out low-confidence frames, and smooth the trajectory with a 5-frame rolling median. The resulting signal is encoded as a 2-channel input comprising relative pitch and raw confidence, with the pitch channel tonic normalized and scaled into octaves to emphasize relative melodic patterns over absolute pitch; uniform pitch-shift augmentation is applied during training. The Conv1DGRU model processes this 1-D input with three Conv1D blocks followed by a deeper three-layer Bi-GRU, since pitch contours carry far less local information than spectrogram inputs and benefit from longer-range temporal modeling~\cite{kum2019joint}.

\begin{table*}[t]
\centering
\small
\setlength{\tabcolsep}{4pt}
\begin{tabular}{l|ccccc|ccccc}
\toprule
& \multicolumn{5}{c|}{\textbf{Daemok-Shared Split}} 
& \multicolumn{5}{c}{\textbf{Work-level Split}} \\
\cmidrule(lr){2-6} \cmidrule(lr){7-11}
Modality 
& Masked & Ujo & Gmj & Anr & Cj
& Masked & Ujo & Gmj & Anr & Cj \\
\midrule
Mel-Original & \textbf{0.871} & 0.777 & \textbf{0.901} & \textbf{0.975} & \textbf{0.831}
             & \textbf{0.785} & \textbf{0.734} & 0.872 & 0.962 & \textbf{0.571} \\
Mel-Sep      & 0.821 & \textbf{0.783} & 0.899 & 0.972 & 0.632
             & 0.762 & 0.714 & 0.864 & \textbf{0.968} & 0.504   \\
F0        & 0.743 & 0.667 & 0.838 & 0.960 & 0.508
             & 0.683 & 0.615 & 0.826 & 0.947 & 0.343 \\
MIDI         & 0.689 & 0.664 & 0.774 & 0.898 & 0.419
             & 0.626 & 0.608 & 0.785 & 0.880 & 0.233 \\
CultureMERT  & 0.696 & 0.538 & 0.829 & 0.864 & 0.551
             & 0.539 & 0.424 & 0.745 & 0.805 & 0.175 \\
\midrule
Ensemble     & 0.862 & 0.777 & 0.896 & \textbf{0.975} & 0.801
             & 0.748 & 0.730 & \textbf{0.873} & 0.965 & 0.428 \\
\bottomrule
\end{tabular}
\caption{Per-representation and ensemble results under two split strategies.}
\label{tab:main_results}
\end{table*}

\subsection{CultureMERT (CMERTClassifier)}
Universal music representations from self-supervised models such as MERT~\cite{li2023mert} have been widely adopted for downstream music classification~\cite{yuan2023marble}. To leverage such representations for culturally specific repertoire, we employ CultureMERT~\cite{kanatas2025culturemert}, a 95M-parameter model continually pre-trained on multi-cultural music corpora. Raw mono waveforms resampled to 24\,kHz are passed through CultureMERT's convolutional front-end, producing 768-dimensional frame embeddings at 75\,fps. We freeze the convolutional feature extractor and all Transformer encoder layers except the one used for adaptation, fine-tuning only the final encoder layer together with a two-layer MLP head. We selected the adaptation point empirically: in a probing experiment we evaluated the output of every Transformer encoder layer independently, and layer 8 achieved the best validation macro-F1. This partial unfreezing strategy preserves the general acoustic priors acquired during large-scale SSL pretraining while allowing the final layer to be re-purposed from its original masked-prediction objective into a task-specific adapter for mode classification~\cite{howard2018universal, li2023mert}.

\subsection{Ensemble}
Frame-level outputs from each modality are produced at native rates and temporally aligned to a common grid. We combine modality-specific predictions via late fusion: softmax probabilities from individual models are averaged frame-by-frame, 
and the argmax yields the final prediction. The ensemble combines modalities corresponding to distinct theoretical axes of mode: Mel-Sep (holistic/performative), MIDI (scale-theoretic), and PESTO (micro-pitch/performative). This theory-grounded composition allows disagreement patterns across components to admit direct musicological interpretation, which we exploit in our qualitative analysis.

\section{Experiments}\label{sec:mode_classification}
\subsection{Experimental Setup}

\textbf{Training.} 
All models are trained for 10{,}000 iterations with the Adam optimizer, 
using a \texttt{ReduceLROnPlateau} scheduler (factor\,=\,0.5, $\mathrm{lr}_{\min}$\,=\,1\textsc{e}-6) driven by validation macro-F1. We evaluate on the validation set every 200 iterations and retain the checkpoint with the highest validation macro-F1 for test evaluation. All models share a base learning rate of $5{\times}10^{-4}$ except CMERTClassifier, which uses $3{\times}10^{-4}$ to accommodate its pretrained backbone. Per-modality batch sizes and scheduler patience values were selected via validation performance. All experiments were 
conducted on a single NVIDIA GeForce RTX 4090.

\textbf{Loss Function.} 
We train all modalities with Focal Loss~\cite{lin2017focal} (focusing parameter $\gamma = 2$), with class weights $w_c$ inversely proportional to class frequency to address mode imbalance. Frames labeled \emph{Unknown} are excluded from both training and evaluation.

\textbf{Evaluation.} We evaluate on a shared test set of 18 pieces under both the Daemok-Shared and Work-level splits. Our primary metric is masked macro-F1, which is computed only over frames with a defined mode label (excluding the \emph{Unknown} class), and we additionally report per-class frame-wise F1.

\subsection{Results}

Table~\ref{tab:main_results} reports masked macro-F1 (the unweighted mean over the four modes) and per-class frame-wise F1 under the two split strategies.


Across the three well-represented modes (Ujo, Gyemyeonjo, Aniri), the split gap averages only $2.1$--$3.6$ points of F1 for all four signal-derived representations, indicating that our models learn mode-relevant musical features rather than memorizing performer- or recording-specific artifacts, with MIDI and PESTO degrading slightly less than the spectral representations. The larger gap in overall masked macro-F1 ($5.9$--$8.6$ points) is driven almost entirely by Changjo, which macro-F1 weights equally despite its $1.8$\% share of labeled frames.


The class-wise pattern is consistent across representations rather than asymmetric: every representation, on both splits, scores higher on Gyemyeonjo than on Ujo, and Aniri is the easiest mode throughout ($0.805$--$0.975$). Gyemyeonjo and Aniri are also the low-variance classes (ranges of $0.127$ and $0.111$ across representations on the Daemok-Shared split), whereas Ujo and Changjo vary by $0.245$ and $0.412$ --- the choice of representation matters far less for the easy modes than for the hard ones. We read the uniform Ujo deficit as evidence that Ujo is the less strongly marked category, defined largely by the \emph{absence} of the ornamental and micro-pitch cues (\emph{sigimsae}) that positively identify Gyemyeonjo --- which aligns with practitioners' own account of this distinction as subtle and difficult.


CultureMERT differs from the signal-derived representations in generalization rather than raw accuracy. It has the largest split gap in every column ($11.4$/$8.4$/$5.9$/$37.6$ points on Ujo/Gyemyeonjo/Aniri/Changjo, against maxima of $6.9$/$3.5$/$1.8$/$26.0$ elsewhere), and the ranking reverses across splits: it edges MIDI on Daemok-Shared ($0.696$ vs.\ $0.689$) but falls $8.7$ points behind it on Work-level, where it is the weakest representation overall. The class-wise skew points the same way: relative to Mel-Original it loses $31\%$ on Ujo and $34\%$ on Changjo but only $8\%$ and $11\%$ on Gyemyeonjo and Aniri, and Gyemyeonjo F1 ($0.829$) still trails Mel-Original's ($0.901$) despite the model defaulting toward that class, since false positives absorbed from Ujo and Changjo depress its precision.

Source separation degrades performance overall ($0.821$ vs.\ $0.871$ on Daemok-Shared; $0.762$ vs.\ $0.785$ on Work-level), and the loss is entirely localized: Mel-Sep stays within $\pm0.006$ of Mel-Original on Ujo, Gyemyeonjo, and Aniri but drops $19.9$ points on Changjo, which at the macro average's $1/4$ weight accounts for the entire $0.050$ overall gap; the Work-level split shows the same signature, with Changjo falling $6.7$ points against at most $2.0$ for any other mode. Changjo's melodic content shares characteristics with both Ujo and Gyemyeonjo, and its identity as a distinct mode is determined primarily by the presence of drum accompaniment~\cite{jo2014}. Removing the percussion strips the very cue that distinguishes Changjo, leaving only the melodic information the three modes share.

The ensemble should be read against its own components. On Daemok-Shared late fusion yields a real gain --- $0.862$ against $0.821$ for the best component --- driven by Changjo, where it reaches $0.801$ against a best component of $0.632$ while matching its components to within $0.006$ on the other three modes. This complementarity does not survive the work-level split, where the ensemble falls to $0.748$ against Mel-Sep's $0.762$ and loses $7.6$ points on Changjo. Neither configuration surpasses Mel-Original overall ($0.862$ vs.\ $0.871$; $0.748$ vs.\ $0.785$). We therefore present the ensemble not as a higher-accuracy system but as an analytical instrument: its components are motivated \emph{a priori} by pansori mode theory, and their disagreement patterns admit musicological interpretation, which we examine in Section~\ref{sec:analysis}.


\begin{figure}[t]
    \centering
    \includegraphics[width=\columnwidth]{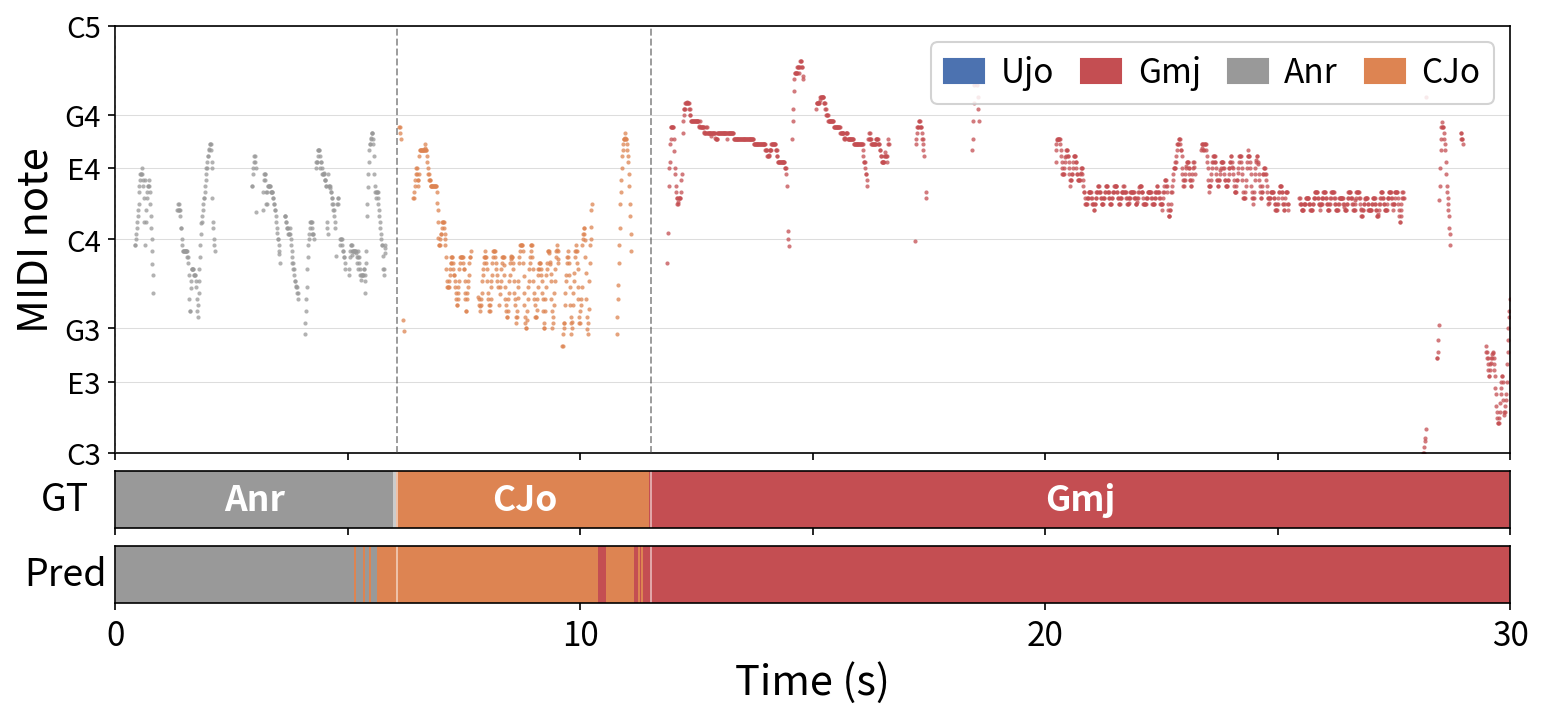}
    \caption{Frame-level F0 contour colored by ensemble prediction, 
ground-truth labels, and ensemble predictions for a 30-second 
passage from Lee Il-ju's \emph{Simcheongga} (``Death of Lady 
Gwak''). The ensemble correctly localizes both mode transitions 
(Aniri$\to$Changjo$\to$Gyemyeonjo).}
    \label{fig:example_prediction}
\end{figure}

\section{Analysis and Discussion}\label{sec:analysis}

To examine in concrete musical terms what each modality has learned, a co-author with formal pansori training qualitatively reviewed the held-out test passages, identifying cases where specific modalities succeeded while others failed, as well as cases where all models converged on a label that contradicts the ground truth. Together these cases reveal that the multi-modal framework operationalizes the multidimensional definition of pansori mode developed in Section~\ref{sec:background}: each modality privileges a different musical dimension, and their 
agreement and disagreement together expose the structure of modal identity.

\subsection{Modality-Specific Musical Lenses}

When examining cases where only a single modality predicted the ground truth correctly, we observe that each input representation captures a distinct dimension of pansori mode.

\textbf{MIDI:}
Passages where only the MIDI input representation model succeeds are characterized by clear interval relationships. Even in fast \textit{jajinmori} passages crowded with vocal noise, MIDI 
identifies Ujo by anchoring on stable repetitions of \emph{mi} and the ascending \emph{do--re--mi} motif, and identifies Gyemyeonjo from the descending perfect fourth 
\emph{la--mi}. The transcription step strips away performative texture and exposes the underlying scale-theoretic skeleton.

\textbf{F0:} 
The Conv1DGRU model succeeds uniquely in passages where mode identity hinges on \textit{sigimsae}. PESTO classifies Gyemyeonjo from the intense \textit{yoseong} (vibrato) on high pitches and the precise semitone bending (do--si), and identifies Ujo from 
its slow, upward-pushing pitch trajectories.

\textbf{Mel:} 
A substantial set of passages is correctly classified only by Mel. Expert review identified that mode identity in these cases rests on vocal timbre and dynamic energy. Mel correctly identifies Ujo from the singer's open vocal production and thick, robust voice coupled with strong accents on opening syllables, and identifies Gyemyeonjo from a narrowed throat and a weaker, descending dynamic envelope. This explains Mel's leading per-class F1 in Table~\ref{tab:main_results} and clarifies its role in the ensemble: in pansori, articulatory texture is itself a defining parameter of mode, and Mel is the only modality that preserves it.

\subsection{Cross-Modal Disagreement and Modal Ambiguity}

The most diagnostic cases are those in which Mel, PESTO, MIDI, and CultureMERT all converge on a prediction that contradicts the annotation. These cases reveal moments where different dimensions of the multidimensional definition pull in opposite directions.

\textit{Pattern~A: scale and tempo overriding sigimsae.} 
Frames annotated \textit{Gyemyeonjo} are systematically misclassified as Ujo when sung in a fast rhythmic pattern over a \emph{do--re--mi} central melody with sustained high pitches. In fast tempos, vocalists naturally reduce heavy sigimsae, 
leaving only the pitch skeleton; stripped of Gyemyeonjo's ornamental signature, the models read the stable, high-energy pitches as Ujo.

\textit{Pattern~B: contour overriding scale.} 
The reverse error appears when annotated Ujo passages contain slow, descending melodic motion through the lower \emph{mi--sol--la} region. Although the underlying scale belongs 
to Ujo, the descending contour and low-energy delivery activate features the models have learned as Gyemyeonjo markers.

The bidirectionality of these errors illustrates the multidimensional definition. The Ujo--Gyemyeonjo distinction does not rest on any single surface feature; when scale, tempo, ornamentation, and articulation contradict one another—a common occurrence in fluid pansori performance—the modalities, each privileging a different dimension, converge on the dimension most strongly activated.

\subsection{Application to Modern Pansori}
A central question for any computational analysis of a living oral tradition is whether models trained on canonical repertoire generalize to contemporary creative practice. We address this by applying our ensemble to three modern (\emph{changjak}) pansori works that do not appear in training: 
\emph{Keunbyeoreun Badae Tteoreojigo} (composer Ahn Hyang-ryeon), \emph{Yi Sun-sin Yeolsa-ga} (composer Jeong Cheol-ho, performed by Cho Sang-hyeon), and \emph{Yi Jun Yeolsa-ga} (composer Jeong Cheol-ho, performed by Sung Chang-sun). Two of these works have been independently analyzed in published musicological studies based on manual score transcription~\cite{chung2026, lee2017yijun}, allowing us to compare the ensemble's frame-level predictions against established score-based mode classification. 

For the two Jeong Cheol-ho works, every mode transition identified in the published analyses is recovered by the ensemble at the corresponding lyric. The agreement extends beyond the boundaries themselves to the underlying musical mechanisms: transitions attributed in the published analyses to characteristic Ujo cues (do--re--mi anchoring, perfect-fourth ascending motion) and Gyemyeonjo cues (\emph{yoseong}, \emph{kkeokneun-eum}) drive the ensemble's predictions in the same direction. For the third work, which has no comparable published analysis, a co-author with formal pansori training confirmed that the ensemble's predicted mode boundaries are consistent with her independent musicological judgment.

Taken together, these results indicate that frame-level computational mode classification, when grounded in theoretically motivated multi-modal representations, agrees with both established musicological theory and expert manual analysis on modern pansori.

\section{Conclusion}
We presented a frame-level pansori mode classification framework on a 46-hour expert-annotated corpus, using complementary representations each tied to a different dimension of pansori mode. Performance is stable under a strict work-level split for the well-represented modes, and the disagreements among representations align with expert musicological analysis, positioning multi-representation modeling as both a practical annotation tool and an empirical lens onto the multidimensional structure of pansori mode.

\section{Acknowledgments}
This work was supported by the National Research Foundation of Korea (NRF) grant funded by the Korea government (MSIT) (RS-2025-00560548).

\bibliography{ISMIRtemplate}

\end{document}